\documentclass[aps,prb,reprint,superscriptaddress]{revtex4-1}
\usepackage[dvipdfmx]{graphicx}
\usepackage{pstricks}
\usepackage{amsmath}
\usepackage{dcolumn}
\usepackage{color}
\usepackage{nicefrac}
\usepackage{bm}
\newcommand{\lixtrue}{Li$_x$(C$_3$N$_2$H$_{10}$)$_{0.37}$FeSe}
\newcommand{\lix}{Li$_x$(C$_3$N$_2$H$_{10}$)$_{0.5}$FeSe}
\newcommand{\dip}{C$_3$N$_2$H$_{10}$}
\usepackage{pdfpages}
\makeatletter
\AtBeginDocument{\let\LS@rot\@undefined}
\makeatother

\begin{document}
%\title{Doping effect on the electronic structure and superconductivity on Li$_x$(C$_3$N$_2$H$_{10}$)$_{0.5}$FeSe}
%\title{Superconducting $T_{\rm c}$ dome as a function of doping in an FeSe intercalate}
%\title{Doping-controlled superconducting dome in FeSe intercalates}
%\title{Interplay of Fermi surface and magnetic interactions generates the superconducting dome in electron doped FeSe intercalates}
\title{Importance of Fermi surface and magnetic interactions for the superconducting dome in electron doped FeSe intercalates}

\author{Makoto Shimizu} 
\affiliation{Department of Physics, Okayama University, Okayama 700-8530, Japan}
\author{Nayuta Takemori} 
\affiliation{Research Institute for Interdisciplinary Science, Okayama University, Okayama 700-8530, Japan}
\author{Daniel Guterding} 
\affiliation{Fachbereich Mathematik, Naturwissenschaften und Datenverarbeitung, Technische Hochschule Mittelhessen, Wilhelm-Leuschner-Stra{\ss}e 13, 61169 Friedberg, Germany}
\author{Harald O. Jeschke}
\affiliation{Research Institute for Interdisciplinary Science, Okayama University, Okayama 700-8530, Japan}

\begin{abstract}
The van-der-Waals gap of iron chalcogenide superconductors can be intercalated with a variety of inorganic and organic compounds that modify the electron doping level of the iron layers. In {\lixtrue}, a dome in the superconducting transition temperature $T_{\rm c}$ has been reported to occur in the doping range of $x=0.06$ to $x=0.68$. We use a combination of density functional theory and spin fluctuation theory to capture the evolution of superconducting transition temperatures theoretically. We clearly demonstrate how the changing electronic structure supports an increasing superconducting $T_{\rm c}$. The suppression of $T_{\rm c}$ at high doping levels can, however, only be understood by analyzing the magnetic tendencies, which evolve from stripe-type at low doping to bicollinear at high doping.
\end{abstract}

% \pacs{
%   71.20.-b, %Electron density of states and band structure of crystalline solids
% }

\maketitle

%Introduction section
{\it Introduction.-} Iron chalcogenide superconductors, while structurally simple, have emerged as one of the most interesting and electronically most complicated of the various families of iron-based superconductors~\cite{Pustovit2016,Boehmer2018,Coldea2018}. FeSe, in particular, is intensely studied due to its nematic phase which occurs in a very large temperature range~\cite{McQueen2009} and due to its peculiar electronic structure that is not correctly captured by any known electronic structure technique~\cite{Yi2019}. After the discovery of superconducting FeSe~\cite{Hsu2008} it was realized that the superconducting transition temperatures can be significantly enhanced if the van der Waals gap between the iron selenium layers is intercalated with alkali metal ions, for example by the ammonia technique~\cite{Scheidt2012,Burrard2013,Sedlmaier2014,Shahi2018} or directly in the structurally complicated $A_x$Fe$_{2-y}$Se$_2$ type of compounds with $A={\rm K}$, Rb, Cs~\cite{Bao2013,Krzton2016}. Further possibilities are completely inorganic lithium hydroxide intercalates~\cite{Sun2015,Lu2015}, and alkali metal intercalates stabilized by organic solvents~\cite{Noji2014}. Not only FeSe but also FeS~\cite{Zhou2017} and FeSe$_{1-x}$Te$_{x}$~\cite{Zheng2016,Li2017} can be intercalated.

At least three tuning parameters have been identified that provide control over the superconducting transition temperature: Pressure can increase the $T_{\rm c}$ of bulk FeSe from 8~K at ambient pressure to 36.7~K~\cite{Medvedev2009}. By electron doping via lithium and ammonia intercalation, $T_{\rm c}$ up to 55~K can be reached~\cite{Shahi2018}. The spacing between FeSe layers can be systematically increased using amines of increasing size, like ethylenediamine (C$_2$H$_8$N$_2$)~\cite{Noji2014,Jin2017}, 2-phenethylamine (C$_3$H$_{11}$N)~\cite{Hatakeda2016}, diaminopropane (C$_3$H$_{10}$N$_2$)~\cite{Jin2017,Sun2018}, hexamethylenediamine (C$_6$H$_{16}$N$_2$)~\cite{Hosono2014,Hosono2016}, and  $T_{\rm c}$ up to 41~K has been observed~\cite{Hosono2016}. 

Tuning parameters can also be combined, for example by pressurizing doped FeSe intercalates~\cite{Sun2012,Izumi2015,Song2016,Sun2018b}. When parameters like pressure or doping, which can be varied continuously, are tuned away from their optimal values, characteristic superconducting domes are observed~\cite{Sun2012,Izumi2015,Song2016,Zhang2017,Sun2018b}.

Here, we are interested in finding microscopic explanations for superconducting domes, also in the hope of discovering ways to manipulate and increase transition temperatures. Theoretically, superconductivity in iron-based materials has been tackled from a weak coupling perspective, describing Fermi surface instabilities that lead to superconductivity via spin flucutations~\cite{Mazin2008,Kuroki2008,Hirschfeld2011}, and from a strong coupling picture by considering these materials as doped Mott insulators, where superconductivity is obtained from approximatively solved $t$-$J$ models~\cite{Lee2006,Si2008,Davis2013}.
However, capturing superconducting $T_{\rm c}$ trends continues to be a theoretical challenge~\cite{Suzuki2014}.
Within a spin fluctuation scenario, the $T_{\rm c}$ increase with doping has been explained for lithium ammonia intercalated FeSe~\cite{Guterding2015}, and for FeS, a pressure induced double dome was captured~\cite{Shimizu2018}. The dispute between itinerant and localized electron pictures is ongoing, and while some iron-based superconductor families, like 1111, are more itinerant, others, like hole doped 122 and iron chalcogenides are more localized.  In fact, superconductivity can be shown to depend both on Fermi surface shape and antiferromagnetic exchange~\cite{Hu2012}. Even in the absence of a magnetic ground state, the nature of the magnetic exchange has important implications for nematicity and superconductivity in FeSe~\cite{Glasbrenner2015} and for the differences between iron pnictides and germanides~\cite{Guterding2017}.

In this Rapid Communication, we will focus on the lithium diaminopropane intercalated material {\lixtrue} where charge doping has been controlled experimentally in a wide range $0.06\le x\le 0.68$~\cite{Sun2018}. We will work both in the itinerant scenario by applying spin fluctuation theory and in the localized picture by working out the consequences of the Heisenberg interactions. We will show that the underdoped side of the superconducting dome can be well understood based on Fermi surfaces, but in order to explain the decreasing $T_{\rm c}$ at high doping, we need to evoke the high energy argument that, with increasing carrier density, magnetic interactions mutate from FeSe-like to FeTe-like. 

%Here, we will show that the decisive factor which explains deteriorating superconductivity for the overdoped FeSe intercalate is the loss of stripe-type magnetic interactions.

\begin{figure}[t]
  \includegraphics[width=\linewidth]{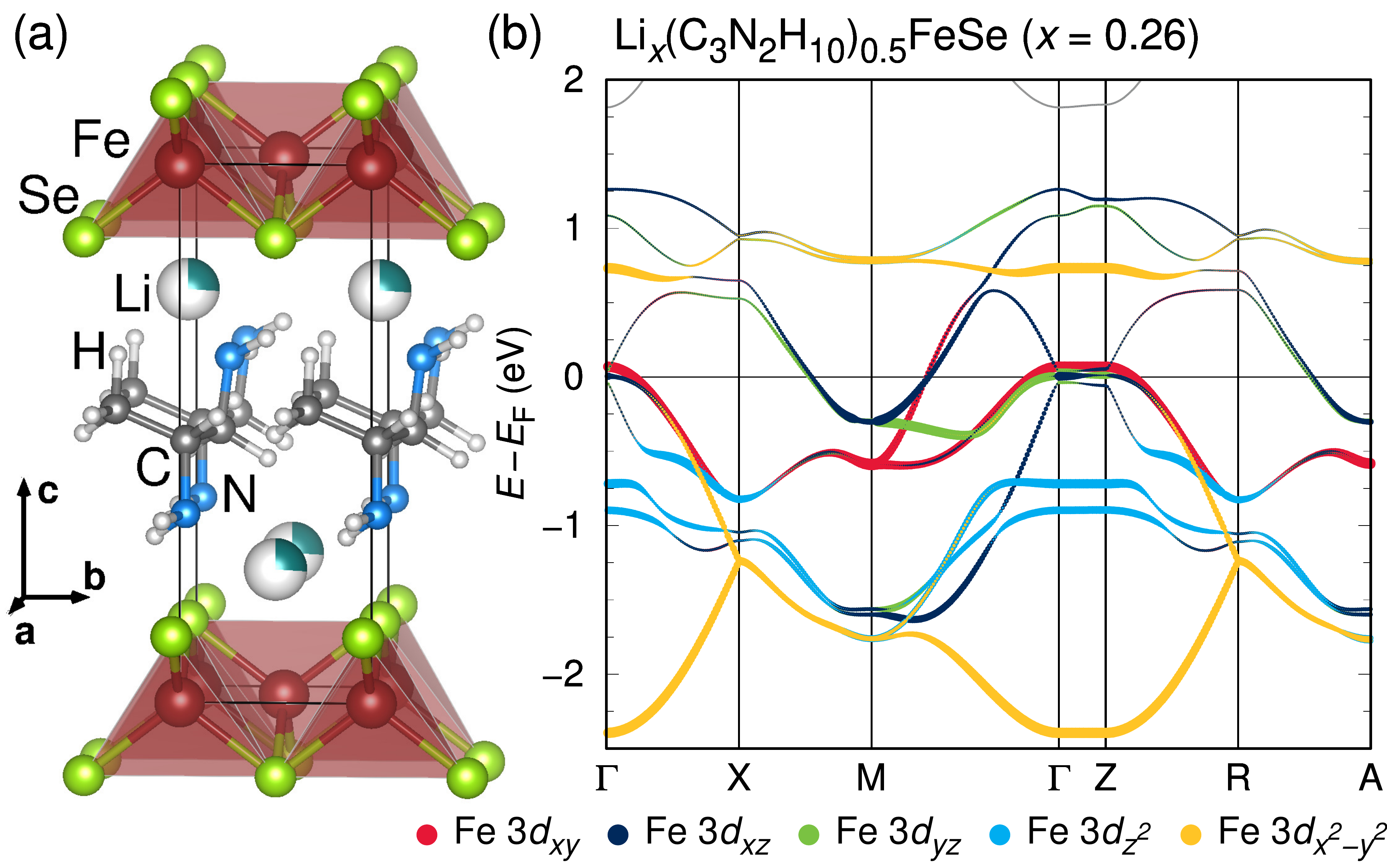}
  \caption{(a) Crystal structure of {\lix}; the Li position is only {26\%} occupied, indicated by partial coloring. (b) Corresponding bands near the Fermi level $E_{\rm F}$ with Fe $3d$ orbital weights.}
  \label{fig:crystal_bweights}
\end{figure}

{\it Methods.-} We study the electronic structure of {\lix} as a function of doping by performing a series of density functional theory (DFT) calculations within the full-potential local orbital (FPLO)~\cite{Koepernik1999} basis, using the generalized gradient approximation (GGA) exchange correlation functional~\cite{Perdew1996} and $24\times24\times24$ $k$-mesh. We construct tight-binding models including all ten Fe $3d$ orbitals using projective Wannier functions~\cite{Eschrig2009}. 
Using the glide reflection unfolding technique~\cite{Tomic2014}, we turn the ten-band into a five-band model. We calculate the noninteracting susceptibility $\chi_0$ on a  $50\times50\times10$ $q$-mesh. We employ the random phase approximation (RPA) to calculate spin and charge susceptibilities and solve a gap equation to obtain pairing symmetry $\Delta$ and eigenvalue $\lambda$~\cite{Graser2009,Guterding2017b}. On the other hand, we use the energy mapping method to measure the exchange interactions of {\lix}~\cite{Guterding2017,Jeschke2019,Mazin2019} and iterative minimization to find ground state spin configurations for the calculated Hamiltonians~\cite{Lapa2012,Supplement}.

\begin{figure}[ht]
  \includegraphics[width=\linewidth]{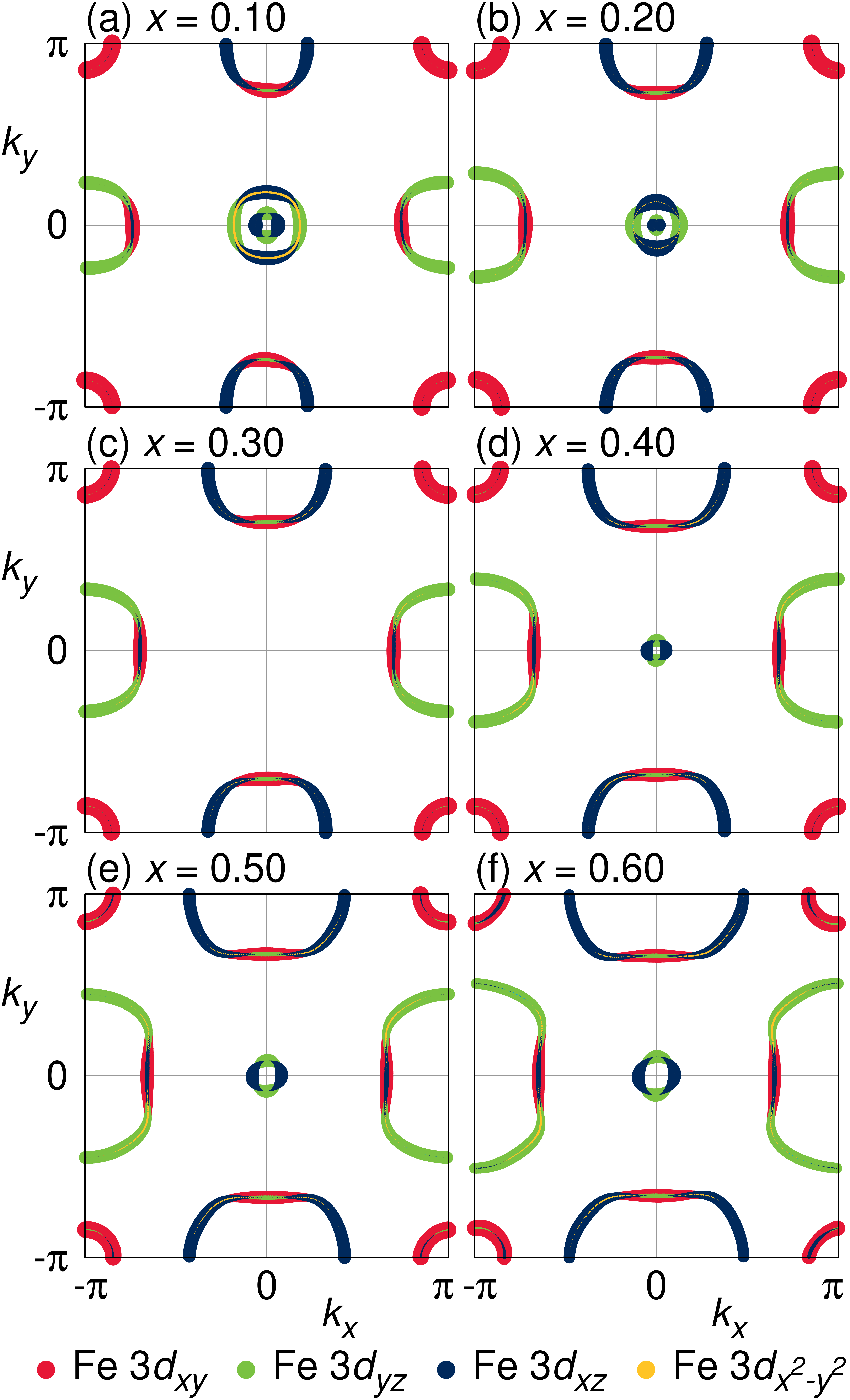}
  \caption{Evolution of the Fermi surface of {\lix} at $k_z=0$ with doping. The Fe $3d$ orbital weights are indicated by color. Orbital weight of Fe $3d_{z^2}$ at the Fermi level is nearly negligible.}
  \label{fig:fs2d}
\end{figure}

\begin{figure}[ht]
  \includegraphics[width=\linewidth]{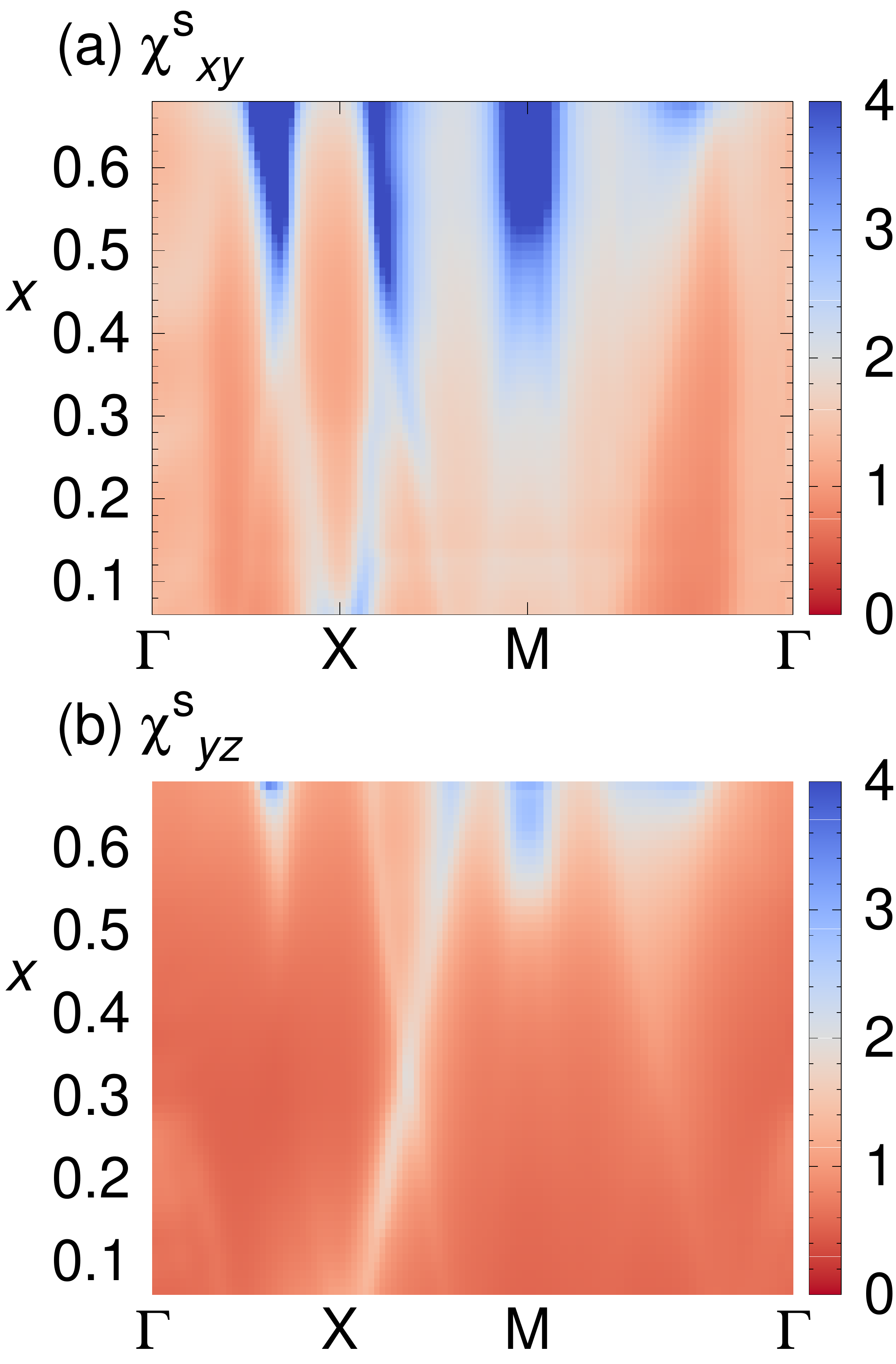}
  \caption{Diagonal elements of the spin susceptibility as a function of momentum vector and doping.}
  \label{fig:spinsuscep}
\end{figure}

\begin{figure}[ht]
 \includegraphics[width=\linewidth]{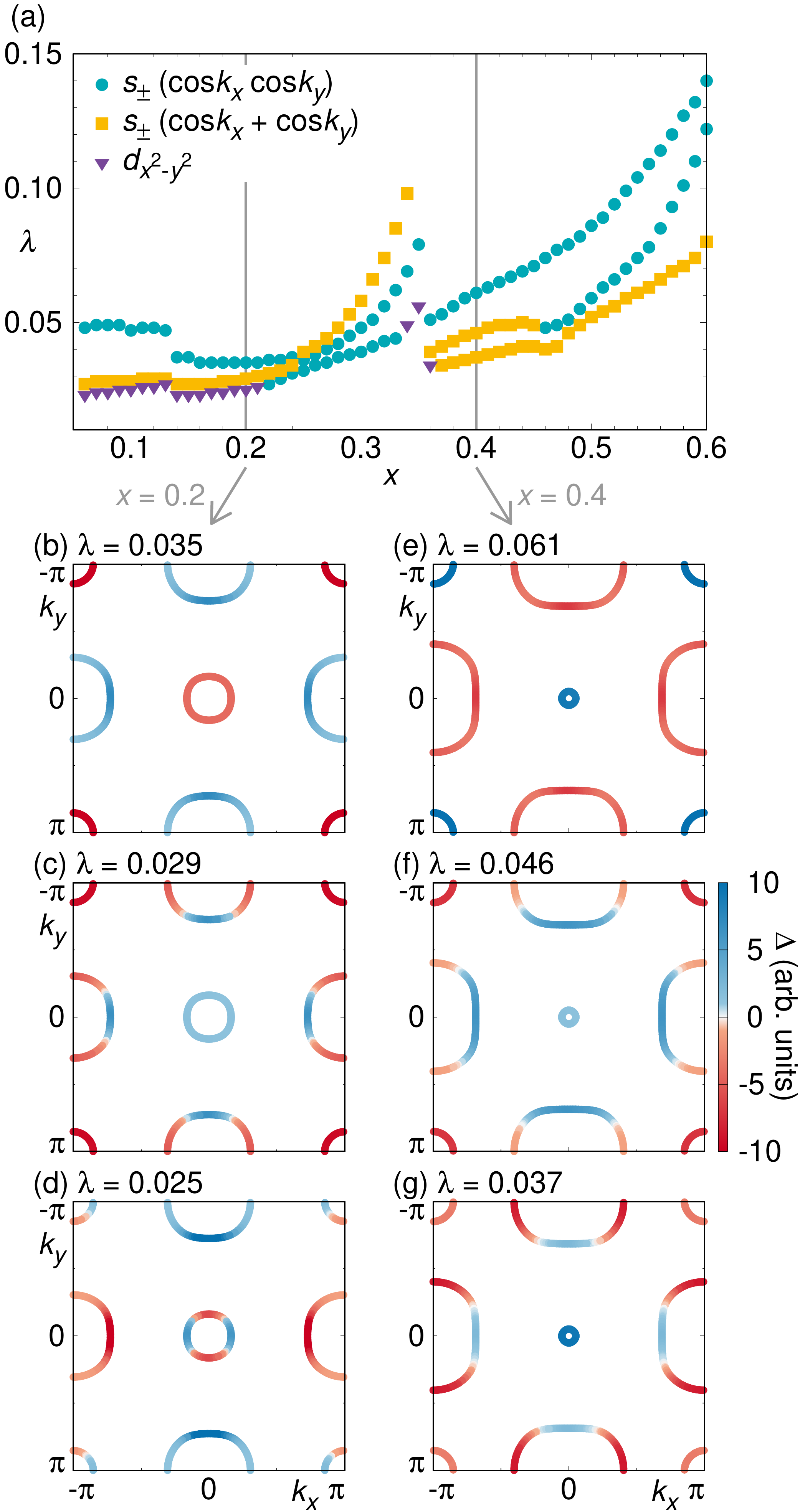}
 \caption{(a) Leading eigenvalues $\lambda$ of the gap equation as a function of doping level $x$. A jump is visible between $x=0.26$ and $x=0.27$, where both of the hole pockets around $\Gamma$ disappear due to a Lifshitz transition. (b)-(d) Leading gap functions on the Fermi surface at doping $x=0.20$. (e)-(b) The same at doping $x=0.40$. All of the results are obtained by calculations without {\dip}.}
 \label{fig:lambda}
\end{figure}

\begin{figure}[ht]
 \includegraphics[width=\linewidth]{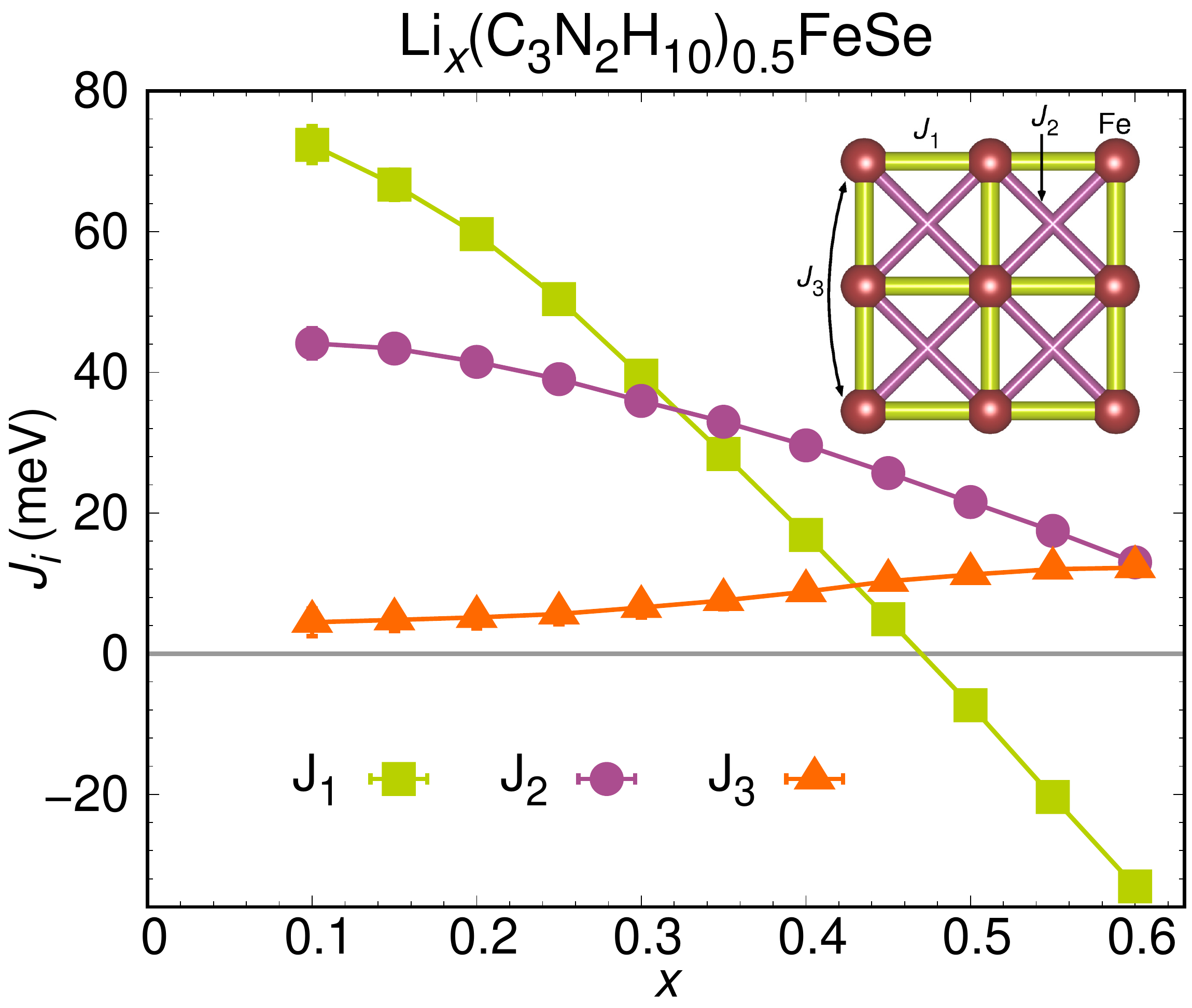}
 \caption{Evolution of three in-plane exchange couplings of {\lix} as a function of doping $x$, obtained from the energy mapping technique. The inset shows the three exchange paths on the iron square lattice.}
 \label{fig:exchange}
\end{figure}

{\it Results.-} We first generate a sequence of crystal structures for {\lixtrue} based on the structural data determined by Sun~{\it et al.}~\cite{Sun2018}. We smoothly interpolate the structural data for the lattice parameters and the FeSe layer (see Ref.~\onlinecite{Supplement} for details, in particular Fig.~S1) and simplify the disordered diaminopropane molecules to a single molecule per two FeSe, yielding the composition {\lix} shown in Figure~\ref{fig:crystal_bweights}~(a). We mount the {\dip} molecule with fixed geometry into the entire structure sequence, since bond lengths and angles of a neutral molecule should not be affected by the charging and contraction of the FeSe layer (see Ref.~\onlinecite{Supplement} for further details). An example of the DFT electronic structure with Fe $3d$ weights is shown in Figure~\ref{fig:crystal_bweights}~(b). Calculations for the complete sequence of structures without diaminopropane molecule, in $P4/mnn$ symmetry, yield similar results~\cite{Supplement}.

We first quantify the charge doping by integration of Fe $3d$ total and partial electron densities. We find that the charge $-xe$ provided by $Li_x^+$ is indeed fully doping the Fe plane; Fe charge evolves almost linearly from $n_{3d}^{\rm total}=6.06$ to 6.65 in the interval $0.06<x<0.68$ (see Ref.~\onlinecite{Supplement}, Fig.~S2).%\ref{fig:num_electrons}).
The orbital with strongest charge increase is $d_{xy}$, followed by $d_{xz}/d_{yz}$.

The decisive factor in spin fluctuation theory is Fermi surfaces. We show in Figure~\ref{fig:fs2d} the unfolded one iron Fermi surfaces of {\lix} for six doping levels (see Fig.~S3 of Ref.~\onlinecite{Supplement} for the same without {\dip} molecules). First of all, we note that the overall number and size of Fermi surface pockets are in good agreement with those observed by angle resolved photoemission in the related FeSe intercalates (Li$_{0.84}$Fe$_{0.16}$)OHFe$_{0.98}$Se and (Tl,Rb)$_x$Fe$_{2-y}$Se$_2$ (see Fig.~1 of Ref.~\onlinecite{Zhao2016}). Even though we do not know the precise doping levels of those compounds, the overall similarity in the Fermi surfaces gives us confidence that for {\lix}, we can avoid the difficulties of bulk FeSe where size and symmetry of Fermi surfaces are captured neither by DFT nor its extensions. Note, also, that DFT+DMFT, as applied in Ref.~\onlinecite{Sun2018}, has only a small effect on the Fermi surface of {\lix}. Thus, the DFT Fermi surfaces are a reasonable starting point for the description of the material. We would like to stress that we take advantage of the fact that in contrast to bulk FeSe where superconductivity develops out of a paramagnetic nematic phase with strong fluctuating magnetic moments which is not fully captured by DFT or DFT+DMFT, magnetic tendencies of FeSe intercalates are more conventional and therefore amenable to the electronic structure methods we employ here.

The dominant feature in the Fermi surface evolution with doping is the strong growth of the large angular electron pockets around the $X$ and $Y$ points in the one iron Brillouin zone (BZ). They have combined $d_{xy}$ and $d_{xz}/d_{yz}$ character. The hole pocket with $d_{xy}$ character at the $M$ point in the one iron BZ is rather insensitive to doping. Finally, several Lifshitz transitions occur around $\Gamma$ with doping, as expected for an iron-based superconductor and previously noted in Ref.~\onlinecite{Sun2018}. In a doping range of $0.24<x<0.30$, an inner hole pocket disappears, an outer hole pocket disappears, and an electron pocket appears; all these are of $d_{xz}/d_{yz}$ character.
The appearance of an electron pocket at $\Gamma$ with doping has also been observed with ARPES on a potassium coated FeSe monolayer~\cite{Shi2017}. Without diaminopropane molecule, the Lifshitz transitions are spread over a range $0.17<x<0.38$, in reasonable agreement with the DFT calculations of Ref.~\onlinecite{Sun2018}. 

The main effect on the density of states at the Fermi level $N(E_{\rm F})$ (shown in Ref.~\onlinecite{Supplement}, Fig.~S4) %the Appendix Figure~\ref{fig:nef})
is a 30{\%} reduction around $x=0.27$ (or $x=0.35$ in the calculation without {\dip}), which results from the disappearance of the two hole pockets; the appearance of the small electron pocket and its growth with doping helps $N(E_{\rm F})$ to recover. 

In the series of calculations without {\dip} molecule, we observe a doping range $0.55 < x < 0.62$, where the Fermi level is pinned to a band crossing. Topological properties of iron-based superconductors have recently been discussed intensively~\cite{Wu2016,Xu2016,Zhang2018,Wang2018}. The presence of diaminopropane molecules in a fixed position as shown in Figure~\ref{fig:crystal_bweights}~(a) breaks the $P4nmm$ symmetry of Li$_x$FeSe and leads to the opening of a small gap in the band crossing. The high symmetry may be restored on average due to the statistical distribution of molecules in the high symmetry of the $P42_12$ space group observed experimentally. Further theory and experiments are required to determine the nature of {\lix} in the doping range $0.55 < x < 0.62$. As the unfolding does not capture the $N(E_{\rm F})$ suppression in this doping range, %(see Appendix Figure~\ref{fig:nef}),
we exclude it from our present discussion.

We now proceed to the spin fluctuation analysis by calculating the RPA spin and charge susceptibilites. As a representative example, we show in Figure~\ref{fig:spinsuscep} the diagonal components $\chi^{\rm s}_{xy}$ and $\chi^{\rm s}_{yz}$ for the structures with {\dip} molecule. 
At low doping, the diagonal element $\chi^{\rm s}_{xy}$ (Figure~\ref{fig:spinsuscep}(a)) is dominated by a peak close to $X$ and a minor peak close to $M$. These correspond to a nesting vector, $\bm{Q}=(\pi+\delta_x, 0+\delta_y, q_z)$ connecting the electron pocket around $X/Y$ and the hole pocket around $M$ point, and a nesting vector $\bm{Q}=(\pi+\delta_x,\pi+\delta_y,q_z)$ connecting the electron pocket around $X$ and that around $Y$. Furthermore, $\chi^{\rm s}_{yz}$ has a peak around $X$, which corresponds to a nesting vector $\bm{Q}=(\pi+\delta_x,0+\delta_y,q_z)$ connecting the hole pocket around $\Gamma$ and the electron pocket around $X$. The nesting vectors change with $\delta_x$, $\delta_y$ because the electron pockets around $X/Y$ grow. In both cases, the susceptibility peak formerly close to $X$ progressively evolves towards $\bm{Q}=(\pi,\pi/2,q_z)$ with increasing doping. This tendency is further accentuated by the Lifshitz transition in the range $0.24<x<0.30$, which also rotates the orbital weights on the $d_{xz}/d_{yz}$ hole pockets around $\Gamma$ by 90 degrees.

Overall, we see an increase in both susceptibilities with doping. This is in line the upward trend of $N(E_{\rm F})$ which is interrupted by the loss of hole pockets around $x=0.27$ and by the band crossing pinned to the Fermi level for $0.55 < x < 0.62$. In agreement with this observation, the trend of the eigenvalue $\lambda$ of the linearized gap equation as shown in Figure~\ref{fig:lambda} is an increase over the entire doping range, with the exception of $x=0.27$ where the Lifshitz transitions cause an abrupt drop of $\lambda$. The leading instabilities are different types of sign changing $s$ wave in the whole region.

Clearly, the itinerant electron analysis of superconductivity in {\lix} presented so far can be used to explain the increasing $T_c$ up to $x=0.37$ observed in Ref.~\onlinecite{Sun2018} (replotted for convenience in Ref.~\onlinecite{Supplement}, Fig.~S5) %\ref{fig:tc})
but not its suppression for higher doping.
As an alternative, we now turn to the localized magnetic moment nature of the Fe $3d$ electrons to search for a mechanism to suppress superconductivity. Figure~\ref{fig:exchange} shows the first three in-plane exchange couplings of {\lix} determined by energy mapping. The inspection of Heisenberg Hamiltonian parameters at this point should not be confused with the assumption of a magnetic ground state; rather, the purpose here is the analysis of the nature of the spin excitations out of which superconductivity is realized.

We find a smooth evolution of exchange couplings over the entire doping range. At small doping, the system starts with all couplings antiferromagnetic, $J_2=0.61J_1$ and $J_3$ nearly negligible. This makes the leading magnetic instability stripe-type AFM~\cite{Supplement}, in perfect agreement with the weak coupling evidence of peaks at $X$ in the spin susceptibility $\chi^{\rm s}$ (Figure~\ref{fig:spinsuscep}). As doping increases, $J_1$ decreases more rapidly than $J_2$ so that at $x=0.33$, $J_1$ and $J_2$ coincide. This evolution strengthens the stripe-type AFM instability. However, the decrease of $J_1$ accelerates until at $x=0.47$, $J_1$ crosses zero and becomes ferromagnetic. Meanwhile, $J_3$ is gradually increasing until at $x=0.6$, it becomes as large as $J_2$. With FM $J_1$ and substantial AFM $J_3$, bicollinear AFM has taken over as leading magnetic instability of {\lix} (see Ref.~\onlinecite{Supplement}, Figs.~S6 and S7). %Figure~\ref{fig:energies}).
Note, that in the strongly doped region, the spin susceptibilities in Figure~\ref{fig:spinsuscep} do show a peak close to $\bm{Q}=(\pi/2,\pi/2,0)$ corresponding to the bicollinear order, but as this is not the dominant instability, the Heisenberg Hamiltonian cannot be understood based on the Fermi surface alone.
%This is consistent with the susceptibilities shown in Figure~\ref{fig:spinsuscep}, which show a peak close to $\bm{Q}=(\pi,\pi/2,q_z)$ in the strongly doped region.

The bicollinear state is the magnetic ground state of FeTe, and as optimally doped FeTe$_{1-x}$Se$_x$ has a $T_{\rm c}$ of only 14.5~K~\cite{Katayama2010}, our calculation provides a strong rationalization of the the falling $T_{\rm c}$ as the magnetism of {\lix} becomes more and more like that of FeTe. We identify the ferromagnetic tendencies of the nearest neighbour exchange as the key factor for determining the $T_{\rm c}$ evolution on the high doping side of the superconducting dome, outplaying all weak coupling effects of growing Fermi surfaces and growing density of states at the Fermi level. 

Note that a similar transition of $J_1$ from AFM to FM was found for a theoretical substitution series between iron pnictides and iron germanides \cite{Guterding2017}, where superconductivity is in fact suppressed in the germanide.
It is an interesting open question if the effect of ferromagnetic nearest neighbour interactions and thus a downward trend in $T_c$ under electron doping could be captured in frequency dependent weak coupling methods like the fluctuation exchange approximation~\cite{Suzuki2014}.

%{\it Discussion}

{\it Conclusions.-} We have analyzed the superconductivity of {\lix} as function of doping in the range $0.06\le x \le 0.68$ both from itinerant electron and localized moment perspectives. We find that on the low doping side $x\le 0.37$, growing Fermi surfaces and densities of states at the Fermi level are the basis for explaining growing $T_{\rm c}$ with spin fluctuation theory. For higher doping level, weak coupling arguments would lead to the erroneous conclusion that $T_{\rm c}$ should continue to grow. However, the progressive destabilization of stripe type AFM fluctuations toward bicollinear antiferromagnetism explains why $T_{\rm c}$ decreases for $x>0.37$. {\lix} turns out to be a perfect demonstration of the fact that iron-based superconductivity can only be understood by fully accounting for itinerant and localized aspects of the Fe $3d$ electrons. Our study highlights that in {\lix}, as in many strongly correlated materials, superconductivity develops out of magnetic interactions that are high energy, not Fermi surface properties, and treating this important class of materials quantitatively calls for further refinement of theoretical methods.

\acknowledgments
The authors would like to thank Hiroaki Kusunose and Igor Mazin for valuable discussions. Part of the computations were carried out at the Supercomputer Center at the Institute for Solid State Physics, the University of Tokyo.

\clearpage


\section*{Supplementary material}

# Importance of Fermi surface and magnetic interactions for the superconducting dome in electron doped FeSe intercalates
## – Supplemental Material –

Makoto Shimizu,[1] Nayuta Takemori,[2] Daniel Guterding,[3] and Harald O. Jeschke[2]

[1]*Department of Physics, Okayama University, Okayama 700-8530, Japan*
[2]*Research Institute for Interdisciplinary Science, Okayama University, Okayama 700-8530, Japan*
[3]*Fachbereich Mathematik, Naturwissenschaften und Datenverarbeitung,
Technische Hochschule Mittelhessen, Wilhelm-Leuschner-Straße 13, 61169 Friedberg, Germany*

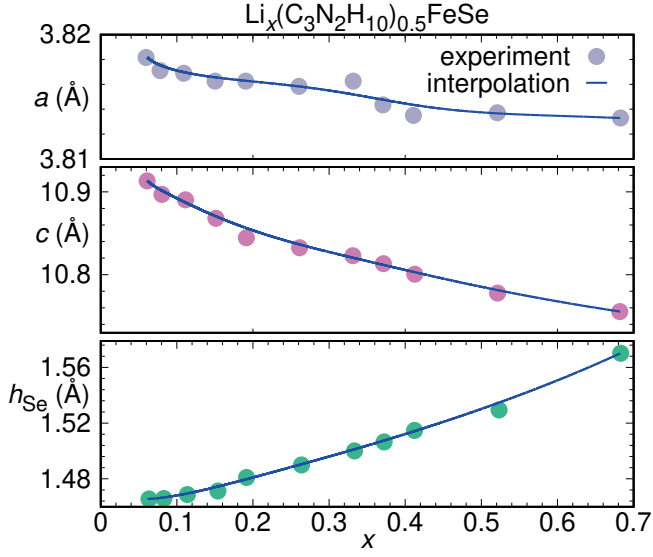

FIG. S1. Structural parameters from Ref. 1 (symbols) and interpolation (lines). Calculations were performed for interpolated structures in steps of $\Delta x = 0.01$.

## I. CRYSTAL STRUCTURES

The study of $\text{Li}_x(\text{C}_3\text{N}_2\text{H}_{10})_{0.37}\text{FeSe}$ requires preparation of model structures as the experimental crystal structure provided in Ref. 1 contains some disorder and is not directly amenable to DFT calculations. First of all, the diaminopropane molecule is placed on an $8g$ Wyckoff position of the $P42_12$ space group with partial occupancy of 0.093, leading to 0.372 $\text{C}_3\text{N}_2\text{H}_{10}$ molecules per FeSe. We simplify this structure by lowering the symmetry to $P1$ and by picking one of the eight symmetry equivalent $\text{C}_3\text{N}_2\text{H}_{10}$ molecules. This leads to the approximate stoichiometry $\text{Li}_x(\text{C}_3\text{N}_2\text{H}_{10})_{0.5}\text{FeSe}$. This structure is shown in Figure 1 of the main text. Note that Sun *et al.* show a very similar simplified structure[1]. The simplification is justified because the diaminopropane molecules are important for fixing the interlayer distance; however, electronically, the highest occupied molecular orbitals are significantly below the Fermi level while the lowest unoccupied molecular orbital is slightly above the Fermi level so that the molecules are not active at the Fermi level. In order to obtain a finely spaced series of crystal structures of $\text{Li}_x(\text{C}_3\text{N}_2\text{H}_{10})_{0.5}\text{FeSe}$ as a function of doping level $x$, we interpolate the structural data provided in Ref. 1 as shown in Figure S1. This guarantees that we do not incur the well-known difficulties of predicting the chalcogenide $z$ position in iron chalcogenide superconductors. Combining the diaminopropane molecule coordinates given for $x = 0.26$ with the lattice parameters of Figure S1 would lead to expanded or compressed molecules as a function of doping. This would be unrealistic as the molecules remain approximately neutral over the whole doping range, and their bonds should be rigid. Therefore, we adapt the diaminopropane molecule Wyckoff positions to the changing lattice parameters, keeping all distances and angles within the molecules constant. We model the doping $x$ by using the virtual crystal approximation for Li, using a nuclear charge between neon and lithium of $Z = 2 + x$.

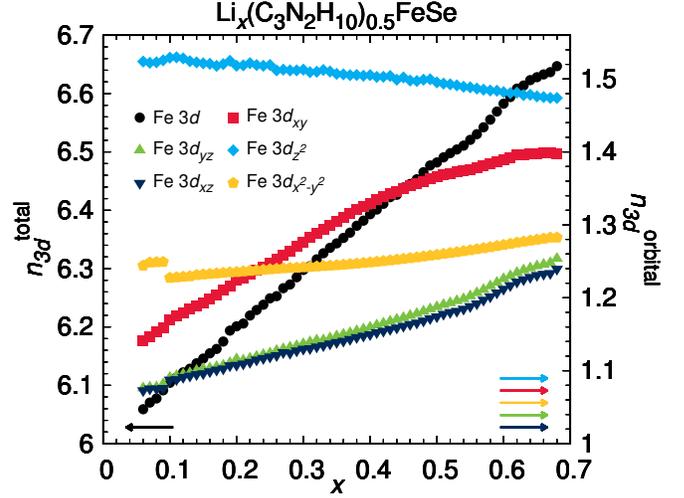

FIG. S2. Evolution of the total and orbital resolved Fe $3d$ occupation number as a function of doping.

## II. ELECTRONIC PROPERTIES

In Figure S3, we show the unfolded one iron Fermi surfaces of $\text{Li}_x\text{FeSe}$ with $\text{C}_3\text{N}_2\text{H}_{10}$ molecules for six doping levels. The corresponding Fermi surfaces of $\text{Li}_x(\text{C}_3\text{N}_2\text{H}_{10})_{0.5}\text{FeSe}$ are shown in the main

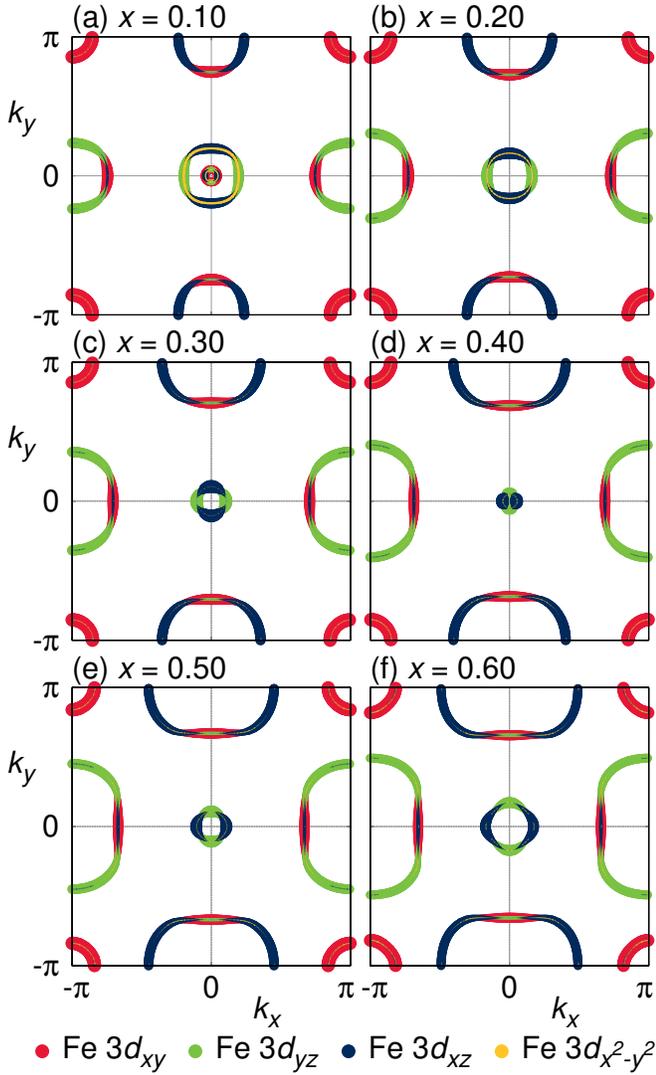

FIG. S3. Evolution of the Fermi surface at $k_z = 0$ with doping for $Li_x FeSe$ without $C_3N_2H_{10}$ molecule. The Fe $3d$ orbital weights are indicated by color. Orbital weight of Fe $3d_{z^2}$ at the Fermi level is nearly negligible.

text. In Figure S2, we show electron densities of $Li_x(C_3N_2H_{10})_{0.5}FeSe$ as a function of doping levels $0.06 \leq x \leq 0.68$. The data are obtained by integrating iron total and orbital resolved densities of states up to the Fermi level.

Figure S4 shows the density of states at the Fermi level $N(E_F)$ with and without $C_3N_2H_{10}$ molecule. $N(E_F)$ of the ten-band tight binding model faithfully represents the DFT result. On the other hand, there is some deviation for the unfolded five-band tight binding model for larger dopings, and in particular in the range $0.55 < x < 0.62$ as discussed in the main text. These deviations stem from the dopants, which lower the crystal symmetry compared to pristine FeSe, so that the unfolding to the five band model becomes only approximate.

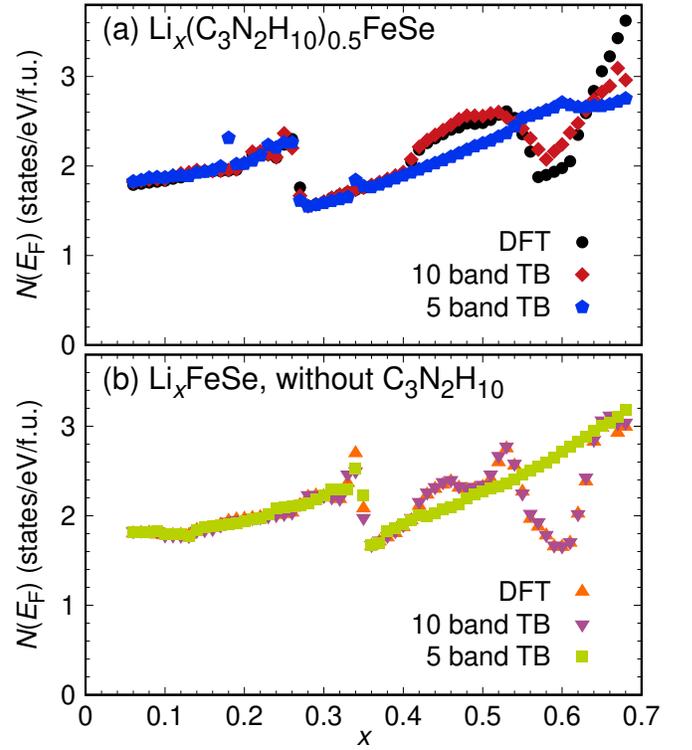

FIG. S4. Density of states at the Fermi level $N(E_F)$ for the two series of structures used in this study. (a) Full structure with diaminopropane molecule, (b) $Li_x FeSe$ with empty van der Waals gap.

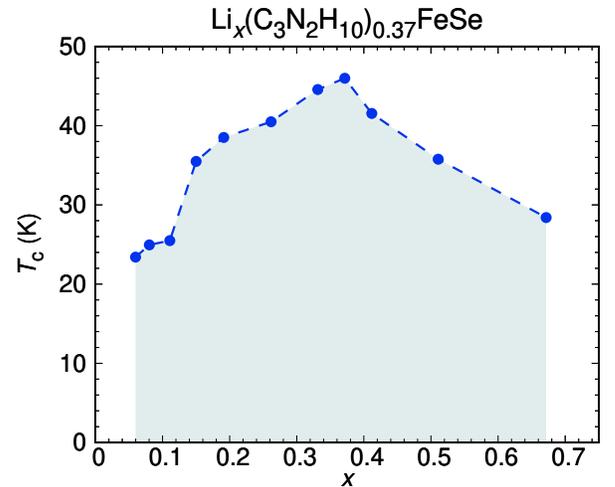

FIG. S5. Superconducting $T_c$ of $Li_x(C_3N_2H_{10})_{0.5}FeSe$ as a function of doping, as reported in Ref. 1.

### III. ENERGY MAPPING

The exchange couplings of $Li_x(C_3N_2H_{10})_{0.5}FeSe$ were determined by the energy mapping technique[2,3]. They are listed in Table S1. $2 \times 2 \times 1$ supercells with $P1$ symmetry contain eight inequivalent iron sites, allowing for 13 spin configurations with different energies. Two exam-

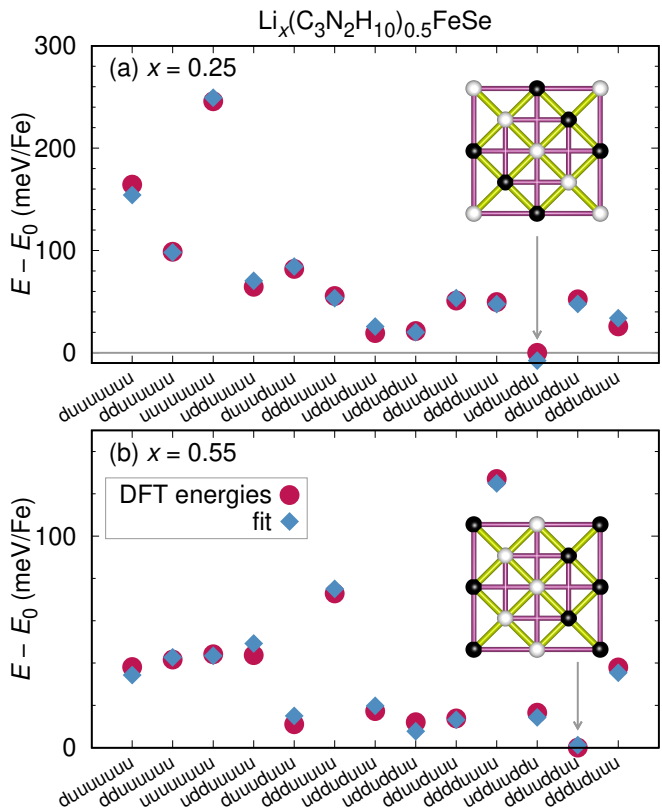

FIG. S6. Two examples for the energy mapping of $Li_x(C_3N_2H_{10})_{0.5}FeSe$ at (a) $x = 0.25$ and (b) $x = 0.55$. In both panels the inset shows an eight iron supercell with the minimum energy spin configuration marked by white for up and black for down. Up to a doping of $x = 0.45$, the stripe-type order is lowest in energy, while from $x = 0.5$, the bi-collinear order is stabilized.

## IV. ITERATIVE MINIMIZATION METHOD

The iterative minimization method allows the efficient determination of ground state spin configurations of classical spin systems[5,6]. We use this method to obtain the ground states of the classical Heisenberg models with the exchange couplings in Table S1. We Fourier transform the obtained spin configurations to obtain the spin structure factor $S(\mathbf{q})$. The spin structure factors of $Li_x(C_3N_2H_{10})_{0.5}FeSe$ for doping values $x = 0.10$ and $x = 0.60$ are shown in Fig. S7. In the doping region of $0.10 \leq x \leq 0.45$, we find peaks at $\mathbf{q} = (\pi, 0)$ or $\mathbf{q} = (0, \pi)$,

TABLE S1. Exchange couplings of $Li_x(C_3N_2H_{10})_{0.5}FeSe$, calculated with GGA and $4 \times 4 \times 4$ $k$ points. The errors shown are only the statistical error arising from the energy mapping.

| $x$ | $J_1$ (meV) | $J_2$ (meV) | $J_3$ (meV) |
| --- | --- | --- | --- |
| 0.1 | 72.3(2.7) | 44.1(2.2) | 4.5(2.0) |
| 0.15 | 66.7(2.2) | 43.4(1.8) | 4.8(1.7) |
| 0.2 | 59.7(2.1) | 41.5(1.7) | 5.1(1.6) |
| 0.25 | 50.4(2.0) | 39.0(1.6) | 5.7(1.5) |
| 0.3 | 39.5(1.9) | 35.9(1.6) | 6.6(1.5) |
| 0.35 | 28.4(1.7) | 33.0(1.4) | 7.6(1.3) |
| 0.4 | 16.9(1.4) | 29.6(1.1) | 8.8(1.0) |
| 0.45 | 4.9(1.1) | 25.7(9) | 10.3(8) |
| 0.5 | -7.3(1.0) | 21.6(8) | 11.2(8) |
| 0.55 | -20.4(1.1) | 17.5(9) | 12.0(9) |
| 0.6 | -33.1(1.1) | 13.0(9) | 12.2(8) |

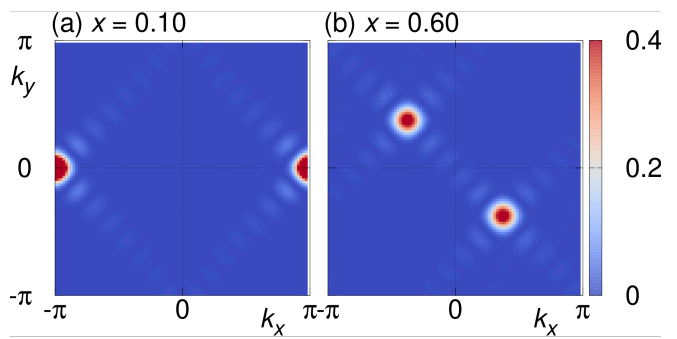

FIG. S7. Spin structure factors $S(\mathbf{q})$ at $x = 0.10$ and $x = 0.60$, obtained by Fourier transforming the ground state spin configuration found with iterative minimization.

ples are shown in Figure S6. Due to the metallic nature of the FeSe intercalate, the fit to the DFT energies is not as good as in magnetic insulators[4] but it is very reasonable as indicated by the small deviations between fitted and DFT energies and by the small error bars shown in Figure 5 of the main text. Note that at large doping, the bicollinear configuration takes over from stripe-type as lowest energy spin configuration (Figure S6).

indicating stripe AFM order as ground states. On the other hand, for $0.50 \leq x \leq 0.60$, as soon as $J_1$ turns ferromagnetic, we find peaks at $\mathbf{Q} = (q, q)$ where $q \sim 0.4\pi$. As it is well known that FeSe also has a substantial biquadratic coupling[7] which we do not re-determine here, the highly doped $Li_x(C_3N_2H_{10})_{0.5}FeSe$ samples will actually be in the bicollinear phase with $\mathbf{q} = (\pi/2, \pi/2)$ and magnetically resemble FeTe.

---



\end{document}